\documentclass{article}
\usepackage{graphicx} 
 
 \usepackage{setspace}
\title{Against Self-Location  }
\author{Emily Adlam  \thanks{Philosophy Department and Institute for Quantum Studies, Chapman University, Orange, CA92866, USA \texttt{eadlam90@gmail.com} }}

     \usepackage[top=2.5cm, bottom=2.5cm, left=2.5cm, right=2.5cm]{geometry}
 \usepackage{setspace}
\begin{document}

\maketitle

\begin{abstract}

I distinguish between pure self-locating credences and superficially self-locating credences, and argue that there is never any rationally compelling way to assign pure self-locating credences. I first argue that from a practical point of view, pure self-locating credences simply encode our pragmatic goals, and thus pragmatic rationality does not dictate how they must be set. I then use considerations motivated by Bertrand's paradox to argue that the indifference principle and other popular constraints on self-locating credences fail to be a priori principles of epistemic rationality, and I critique some approaches to deriving self-locating credences based on analogies to non-self-locating cases. Finally, I consider the implications of this conclusion for various applications of self-locating probabilities in scientific contexts, arguing that it may undermine certain kinds of reasoning about multiverses, the simulation hypothesis, and Boltzmann brains.

\end{abstract}
 
\section{Introduction}

 Self-locating   credences are used in a wide variety of contexts in physics and philosophy. For example, they play a central part in reasoning pertaining to the cosmological multiverse\cite{Benetreau-Dupin2015-BENPRI}, the Everett interpretation\cite{10.1093/bjps/axw004,pittphilsci15195}, the simulation hypothesis\cite{Fallis2023-FALSAS-2,2ac867a4-6887-3270-8211-53d9c0e24445}, the arrow of time\cite{Chen2021-CHETAA-8}, Boltzmann brain scenarios\cite{BuilesForthcoming-BUICIA}, and so on. These applications   presuppose that there exist   certain privileged assignations of self-locating credences which we can use in scientific reasoning in much the same way as we would use ordinary non-self-locating credences or probabilities. 

However, in this article I  argue that pure self-locating credences  are not sufficiently objective to bear the weight that is placed upon them in these kinds of scenarios. Of course, it is well-recognised that self-locating credences are not as objective as `objective chances' and other  kinds of probabilities employed in science: Bostrom tells us they are `\emph{not physical chances but subjective credences}'\cite{Bostrom2002-BOSABO-2}. However, in the literature on   self-locating credences it is clear that they  are not regarded as `subjective' in the most radical subjective Bayesian sense, which would entail  they are constrained only by the requirement of probabilistic consistency. Rather, it is assumed that there are  rationally compelling ways to assign pure self-locating credences, and indeed much effort has been expended on determining the correct assignment in various problem cases\cite{Fallis2023-FALSAS-2,Elga2000-ELGSBA,Elga2004-ELGDDE,Monton2002-MONSBA,Weintraub2004-WEISBA}. That is the position I wish to criticize in this article. I will argue that pure self-locating credences  are `subjective' in the sense that they are not rationally constrained by  anything at all, except possibly the requirement of probabilistic consistency. And  I will argue that credences which are `subjective' in this strong sense are largely not able to support substantive scientific conclusions.

I begin in   section \ref{superficial} by distinguishing between `superficial' and `pure' self-locating credences.   In  section \ref{betting} I will argue that assignations of pure self-locating credences cannot be rationally compelling with respect to pragmatic rationality, because in practical scenarios such as betting,  an assignation of pure self-locating credences simply encodes one's practical goals. In section \ref{analogies} I will argue that assignations of pure self-locating credences  are not rationally compelling with respect to epistemic rationality either, since the `Principle of Indifference' and other such principles are not a priori principles of epistemic rationality. In section \ref{analogies2} I will argue that the analogical strategy sometimes employed to argue for certain assignations of self-locating credences is undermined by some key disanalogies. Finally in  section \ref{applications} I will discuss various scientific applications of pure self-locating credences, assessing the extent to which these applications are appropriate if there are no rationally compelling ways to assign pure self-locating credences.  
 
It should be noted that as a corollary of the view that there is no rationally compelling  way to assign pure self-locating credences, it seems natural to say that there is also no rationally compelling way to update pure self-locating credences; since any probabilistically consistent credences are rational, any updating strategy is permissible, provided that the resulting credences are still probabilistically consistent. On the other hand, there are certainly still rational constraints on how we should update superficially self-locating credences, and it is possible that much of the scholarship on the problem of updating self-locating beliefs could be interpreted  as pertaining to the superficial cases, rather than the pure cases. In this article I will not be able to address the issue of belief-updating in depth, but this would be an interesting topic to explore in future work.

\section{Pure versus Superficially Self-Locating Credences \label{superficial}}

It will be important in this article to distinguish pure  self-locating uncertainty, and the associated pure self-locating credences, from a more superficial kind of self-locating uncertainty. The distinction is most easily expressed in the framework of Lewisian possible worlds, using the terminology of a `centered world,'\cite{Lewis1979-LEWADD} to refer to a pair consisting of a possible world together with a   `center' within that world, which might be a place or time or a certain physically embodied observer\footnote{For clarity, note that in this article an entire `multiverse' is understood to be a single `possible world,' because different universes in a multiverse are usually understood to be causally connectible or to have joint common causes. Thus credences to find oneself in one universe or another within the multiverse are pure self-locating credences.}. Using this terminology, I will henceforth take it that pure self-locating (PSL) uncertainty refers to cases in which an observer is uncertain about which centered world they are in, out of a reference class of centered worlds which all belong to the same possible world. Whereas superficially self-locating (SSL) uncertainty refers to cases in which an observer is uncertain about which centered world they are in, out of a reference class of centered worlds  which all belong to different possible worlds. There also exist scenarios involving mixtures of SSL and PSL uncertainty, in which some of the centered worlds  belong to different possible worlds and some of them belong to the same possible world - this is the case in the famous Sleeping Beauty problem\cite{Elga2000-ELGSBA} - but I will not deal with these cases in this article\footnote{However,  I think one consequence of the main thesis of this article is that the Double Halfer position\cite{Meacham2008-MEASBA} is the correct response to the Sleeping Beauty problem.}. 

For an example of superficially self-locating uncertainty, suppose that on days when I do not set an alarm, I do not know what time it is when I wake up. So after I have woken but before I have consulted a clock, I am in a state of uncertainty, and I may assign credences to various times that it might be.   In a sense this is self-locating uncertainty, since it is about `when' I am located. But it is  only superficially  self-locating uncertainty, since in every possible world there is exactly one  time at which I actually wake up on any given morning, so all the different times that I assign credences to are in fact associated with  centered worlds belonging to different possible worlds.  Usually in  cases of SSL uncertainty the actual `location' is  determined by a specific physical process, so for example in this case my actual location in time is determined by the set of biological processes which result in me waking up at a certain time. Hence in cases of SSL uncertainty like this one, it seems natural to say that we should as far as possible assign credences  which appropriately reflect relevant features of the underlying process which determines the `location.' 

For an example of pure self-locating uncertainty, consider Elga's `Dr Evil' scenario\cite{Elga2004-ELGDDE}, in which a person who currently believes himself to be Dr Evil receives a credible message telling him that a subjectively identical duplicate of Dr Evil has been created. This person  is now in a state of self-locating uncertainty, because he does not know whether he is Dr Evil or the duplicate. And since Dr Evil and the duplicate exist within the same possible world, the two possibilities correspond to centered worlds within the same possible world, so this is pure self-locating uncertainty. A characteristic feature of  PSL uncertainty is that there is not any physical process which determines the actual `location,' so for example in this case there is no  physical process by which a certain indexically individuated person is `dropped' into  Dr Evil or the duplicate: there are physical facts about the existence of Dr Evil, and physical facts about the existence of his duplicate, but then there are no further physical facts. Thus wherever PSL credences come from, they cannot simply reflect features of the physical process which determines the `location,' because there is no such process.

Now,  of course  no real agent knows exactly which  possible world she is in, so the definition of pure self-locating uncertainty  as pertaining to set of centered worlds all belonging to a single possible world is an unrealizable idealization.  Rather  in realistic cases of PSL uncertainty there is a set $\{ P_1, P_2 ... P_N \}$ of possible worlds to which the agent assigns non-zero credence, where each $P_{i}$ includes a set $\{ C^1_{P_i},  C^2_{P_i}... C^M_{P_i}  \}$ of   centered worlds that she could be located in. For example, in Elga's Dr Evil case the observer is presumably uncertain about many things other than whether or not he is Dr Evil, so there will be a range of possible worlds that he could be in, all containing an individual who can be identified as that world's version of Dr Evil; so if the agent is now given reliable information that he is in fact Dr Evil, he learns that whatever possible world he should happen to be in, he  is located in the centered world centered on that possible world's copy of Dr Evil, but this piece of information doesn't give him any independent information about which possible world he is in.  

In this article, I will largely focus on   the idealized case in which one is simply deliberating over centered worlds all existing in the same possible world, in order to argue  that there is no rationally compelling way to assign credences over these centered worlds. However, I think it follows from this claim that in the more realistic case where we are considering several possible worlds, there may be a rationally compelling way to distribute credences over the whole possible worlds, but after this distribution has been fixed, we are then free to further distribute credences over the centered worlds within each world in any probabilistically consistent way - there is no rationally compelling way to split the credence assigned to a given possible world between the centered worlds associated with that possible world. So my conclusions for the straightforward case also have implications for the more realistic case. I will   discuss the more realistic case further in section \ref{confirmation}.

 \subsection{ Indexical Self-Reference \label{indexical}}

A key characteristic of genuine PSL uncertainty is that it involves scenarios where there are only two possible ways of singling out an  individual observer from the reference class over which we are uncertain. From a first-person point of view, an observer belonging to the reference class can   use  indexical self-reference to identify herself; but from a  third-person point of view, we can identify a specific observer only by specifying which centered world she is  in, or by giving information which is equivalent to this specification. Genuine PSL uncertainty must have this feature, because if we could  first identify an observer by describing some non-indexical feature $F$ of hers and  then ask whether she is in the centered world $X$ or the centered world $Y$, we would be dealing with centered worlds belonging to different possible worlds: one world in which the observer with property $F$ is in centered world $X$, and another world in which the observer with property $F$ is in centered world $Y$. 

Now, this does not mean that all of the observers must be identical - they will in general have different positions in space or time, different physical features, and so on. Rather it means that in order for you to be in a state of pure self-locating uncertainty, it must be the case that all of the externally-describable identifying characteristics of the observers in your reference class have already been fully specified, so the only remaining locus of uncertainty is about which one of these agents you are.  For example, suppose we perform an experiment in which two identical copies of an observer are made and the original observer is destroyed, and then one copy has her hair dyed blue and the other has her hair dyed green.  After the copying but before seeing the colour of their hair, the copies are in a state of PSL uncertainty, and they may assign self-locating credences over  two centered worlds belonging to the same world, respectively centered on  `the copy with blue hair' and `the copy with green hair.' Now, the two observers will have different physical locations after the copying, so one might think that we could give a third-person description of the experiment in which we first identify a copy by their physical location, and then assign non-trivial credences over whether that copy has blue or green hair. But if that is possible, we end up with SSL uncertainty rather than PSL uncertainty, because we are thereby assigning credences over centered worlds belonging to different possible worlds -   one possible world in  which the observer at the far left has blue hair and another possible world in which the observer at the far left has green hair.  Therefore  in order to construct a scenario involving genuine PSL uncertainty we must insist on a description in which all of the physical facts about the relevant centered worlds are completely  fixed - for example,  we might should specify in advance that the the blue hair will be on the far left observer and the green hair will be on the far right observer - so the only remaining question is an indexical one: `Which of these centered worlds am  I myself located in?' 

This implies that we can only have genuine PSL uncertainty in a case where all observers in the refrence class are subjectively identical, i.e. there is no feature of their internal experience which could give them any information about which particular observer they are.  For if some observer were not subjectively identical to the others in the reference class, then we could identify her without saying which centered world she is in by simply specifying the content of her subjective experience, so this would not be PSL uncertainty. For example, in the case above, suppose that exactly one of the hair dyes causes the eyes to sting, so one of the observer will wake up with sore eyes and the other will not. The observer with sore eyes doesn't know which colour of dye causes the stinging, so she is still uncertain about the colour of her hair, but now the two possible centered worlds she is contemplating belong to two different possible worlds here - one worlds in which  the observer with sore eyes has blue hair, and a different world in which the observer with sore eyes has green hair. So although there is still uncertainty for the observer, it is now SSL uncertainty rather than PSL uncertainty.

This points to an important conceptual difference between PSL and SSL cases. In the PSL case we can't assign any non-trivial third-person credences over the set of  centered worlds, because from the third-person standpoint we can only identify observers by saying what centered world they are in, and thus the only relevant propositions  that we can formulate are of the form `the observer in centered world $X$ is in centered world $Y$' -  and of course this proposition will necessarily be assigned credence $1$ if $X = Y$, and $0$ otherwise. So non-trivial PSL credences really  make sense only from a first-person point of view, since we need to define them using indexical self-identification. By contrast, SSL credences can be formulated from a third-person  perspective - for example, in the case where I am uncertain about what time I woke up, exactly the same credences  may be assigned from a third-person perspective, in which case they will be interpreted as credences concerning the duration of time that a human with a specified causal history and biological features will sleep. Thus SSL credences can simply inherit their values from ordinary third-person physical probabilities, but  PSL credences cannot be directly derived from  any ordinary physical probabilities, so they are `subjective' in a much stronger sense than SSL credences.

However, the literature often fails to distinguish clearly between PSL and SSL uncertainty, leading to problematic equivocations. For example, when Bostrom argues that observer-relative  self-locating credences don't require some kind of special non-physical facts, he  imagines a situation which at first appears to be a PSL case with a number of copies of a human brain being made, but then he argues that we can understand these  credences in physical terms, as follows: `\emph{Let Alpha be  the brain that was recently in states   $A_1$, $A_2$ ... $A_n$. The conditional probability of A being labeled `the bookie' given that A is one of two existing brains is greater than the conditional probability of A being the brain labeled `the bookie' given that A is one out of eleven brains}'\cite{Bostrom2002-BOSABO-2}. But if it is possible to identify observers by appeal to their past brain-states in this way, then there is in fact a non-indexical means of identifying an observer without saying which centered world they are in,  and thus we are switching from PSL uncertainty to SSL uncertainty: if we assume that only one of the brains has had this particular series of brain states (which Bostrom's argument seems to require) then there is one possible world in which the brain that has recently been in the states   $A_1$, $A_2$ ... $A_n$ is the bookie, and another in which it is not the bookie, so the relevant  centered worlds now belong to different possible worlds.  Thus there is some equivocation in this argument: Bostrom has successfully argued  that SSL credences can be understood in purely physical terms, but this does nothing to assuage the concern  that PSL credences cannot be understood in this way, since PSL credences are conceptually very different from SSL ones. 

It should be noted that nearly all cases of self-location uncertainty that we encounter in our everyday lives involve only SSL uncertainty. For example, perhaps the most common kind of self-location uncertainty is uncertainty about what time it is, and as in the case described above this can typically be understood as SSL uncertainty   about the time at which some event occurs. Our intuitions around self-location are therefore largely driven by our experience of SSL uncertainty. It is important to keep this in mind, because  many arguments for certain specific assignations of self-locating credences in PSL cases  involve appeals to intuition, and yet those intuitions have largely been developed for SSL cases, not PSL cases. In particular, there is often a rationally compelling way to assign SSL credences, since such credences can simply be rewritten as NSL (non-self-locating) credences, and therefore the transfer of intuitions from SSL cases encountered in everyday life across to PSL cases encountered in scientific contexts helps to create the impression that there must be a rationally compelling way to assign credences  in these  PSL cases as well. But since  PSL uncertainty is conceptually quite different from SSL uncertainty,  we should be very cautious about applying intuitions based on SSL uncertainty to  PSL cases, and therefore we should perhaps be suspicious of the   idea that there exists a rationally compelling way to assign credences in PSL cases.

\section{Self-Locating Credences and Betting \label{betting}}

If there were a rationally compelling way to assign credences in PSL cases, on what grounds  could such an assignation be rational? There is an ongoing debate about whether reasons for belief should be pragmatic, epistemic, or both\cite{BradleyForthcoming-BRARFB}; here I will not take a position on this debate, but will address both possibilities in turn. Let us begin with   pragmatic rationality, i.e. the kind of rationality that concerns how best to achieve our practical ends. That is, once one has  decided on a set of goals, pragmatic rationality prescribes how one ought to act  to achieve those goals, in light of the practical realities to which one is subject. 

Assigning credences may not initially appear to be   a form of action, but we can make a connection to action via decision-theoretic representation theorems \cite{Meacham2011-MEARTA}, which show that actions taken by a rational agent making choices under uncertainty can be modelled as if that agent is maximizing utility with respect to some particular credence assignation and utility function. So we can see how pragmatic rationality may be thought of as constraining  credences:  it may be that in order to achieve a certain goal, it is pragmatically rational to behave as if one is maximizing   utility with respect to some particular utility function and assignation of credences. 

For example, suppose you are placing bets on the outcome of some probabilistic process with a set of possible outcomes $\{ i \}$, and let $W_S(i)$  be  the winnings you will obtain when the outcome labelled $i$ occurs, if you bet in accordance with strategy $S$. Suppose also that your goal is to achieve the greatest possible  winnings over a large number of trials, i.e. you aim to  obtain the highest possible value for the goal quantity $G = \sum_j \sum_i W_S(i) \delta(i, O(j))$ where $O(j)$ represents the outcome of the process on the $j^{th}$ trial, and we sum over all outcomes $i$, and over a large number of trials $j$. Then we may appeal either to empirical tests or theoretical analysis to show that in order to obtain the highest  value for $G$, you should choose a strategy $S$ which maximizes the quantity $ W( \{ p_i \}) =  \sum_i p_i W_S(i)$ with respect to a certain set of values   $\{ p_i \}$ - that is, you should act as if you are maximizing utility with respect to the credences $\{ p_i\}$, with your utility function given by your winnings $W_S(i)$. Thus  we may argue that you are rationally compelled to assign credences proportional to $\{ p_i \}$, or at least, to act as if those are the credences you assign, since acting in accordance with any other assignation will achieve worse outcomes. As argued in ref \cite{Meacham2011-MEARTA}, showing that one ought to behave as if one assigns certain credences is not necessarily the same as showing that one actually ought to assign those particular credences, but I will assume here  that in order to make the case that that a certain credence assignment is in some sense  rationally compelling, it is enough to show that it is pragmatically rational to act as if one assigns these credences  - for after all, giving up this assumption can only make it more difficult to argue that certain PSL credence assignments are  rationally compelling.

In fact, pragmatic arguments like this have  been made in support of certain assignations of PSL credences. For example, in Bostrom's Dungeon thought-experiment, he argues for a certain assignation of PSL credences on the grounds that if the prisoners bet in accordance with these credences `\emph{then 90\% of all prisoners will win their bets; only 10\% will lose}' and later `\emph{a probability of 90\% is the only one which would make it impossible to bet against them in such a way that they were collectively guaranteed to lose money}'\cite{Bostrom2002-BOSABO-2}. Likewise Leslie argues for a certain assignation of PSL credences on the grounds that `\emph{if every emerald-getter in the experiment betted (in accordance with these credences), there would be five thousand losers and only three winners}'\cite{41604a05-62f9-3cbc-8810-764c1d922162}. These arguments aim to show that a certain credence assignment $  \{ p_i \}$ is rationally compelling on the grounds that choosing a strategy which maximizes the quantity  $ W( \{ p_i \}) =  \sum_i p_i W_S(i)$ will yield the highest winnings summed over all centered worlds in some reference class, i.e. such strategies will yield the highest value of the goal quantity  $G = \sum_i   W_S(i)$, where  $W_S(i)$ is the winnings obtained by the observer in the centered world labelled $i$ if they bet in accordance with strategy $S$, and the sum is taken over the complete set of centered worlds $i$ described in the setup of the thought experiment. 

However, there is something a little odd about this approach. For unless I am unusually altruistic,  when I make bets what I care about is maximizing my own winnings: I don't care how much is won by  other observers in some reference class! So why exactly should I be be required to adopt a strategy which aims to achieve the highest possible value of $G = \sum_i   W_S(i)$?

Well, the problem I face in trying  to design a strategy which benefits me specifically is that, if we take it that we are dealing with genuine PSL uncertainty, all of the observers in the reference class are subjectively identical and  thus there's simply no way I can design a strategy which benefits me more than the other observers in the reference class, since I can't know which observer I am. So it may be  tempting to argue that under these circumstances, maximizing winnings over the whole reference class is my only viable option, even if I don't care at all about the other observers in the reference class. However, this is not true. For example, let us repurpose an idea suggested by   Albert in the context of Everettian probabilities\cite{Albert2010-ALBPIT}.  Suppose that I go to sleep and five copies of me are made, and the copies are put into induced comas for one year, while my original body is destroyed. The numbers from the set $\{ 1, 1, 2, 4, 10 \}$ are assigned at random to the copies, and the amount of nutrition provided to each successor per day is proportional to the number assigned to her, so the five successors have widely varying masses when they wake at the end of the experiment. Imagine that upon waking, and before having an opportunity to gain any information about her current mass, each successor is asked to place a bet on the value of the number that she was assigned - again, we are assuming that the copies are subjectively identical and therefore they must all adopt the same approach. So what bet should the copies make? Well, if the goal is to obtain the greatest possible winnings summed over all  five copies then they should all bet `1,' i.e. they should behave as if they are maximizing utility with respect to a credence distribution assigning equal credence to all five copies. But Albert envisions the possibility of an agent who cares more about her successors with greater mass, perhaps on the grounds that `more is better.' And likewise we can imagine the copies in this experiment deciding that they assign greater utility to winnings accrued by a more massive successor, so the goal will be  to maximize winnings over mass instead, i.e. to choose a strategy $S$ which achieves the highest value for the goal quantity $G = \sum_i c_i W_S(i)$, where $c_i =  m_i \ / \sum_i m_i$, with $m_i$ the mass of the copy labelled $i$. Then if we believe the successor assigned the number 10 will end  up with more than twice as much mass as a  successor assigned the number $1$, the best way to maximize $W$ is to have all the copies bet `10.' That is, the copies will now behave as if they are maximizing utility with respect to a set of credences proportional to mass, $\{ c_i\}$, rather than assigning equal credence to all five copies. 
 
For a more realistic example, recall that if  standard statistical mechanics is right, there may be good reasons to think the world contains many more Boltzmann brains than actual people, and thus  there will likely be a large number of Boltzmann brains having experiences subjectively identical to the ones I am having now\cite{BuilesForthcoming-BUICIA,carroll2017boltzmann, Srednicki_2010}. So it has sometimes been argued  that statistical mechanics on its own implies that you should  believe  you are probably a Boltzmann brain rather than a persisting human individual. However, suppose   you are  asked to place a bet on whether or not you are a Boltzmann brain. If your  goal is to obtain the greatest possible winnings summed over all the whole reference class of individuals subjectively identical to you, i.e. to maximize $G = \sum_i W_S(i)$, then you should assign  equal credences over all subjectively identical centered worlds, including both persisting people and Boltzmann brains -   but is that the only plausible goal here? After all, the large majority of Boltzmann brains exist only for a moment, so even if they do win the bet, they will not last for long enough to enjoy their winnings. Thus it would surely be reasonable for you to decide that you don't care how much is won by  Boltzmann brains, in which case you should adopt a strategy which  aims to obtain the greatest possible winnings for persisting individuals only, excluding Boltzmann brains. Then you and your subjectively identical fellows will be  aiming to achieve the highest possible value for the goal quantity $G = \sum_i c_i W_S(i)$, where $c_i$ has the same value for all persisting individuals and zero for Boltzmann brains, and this will  lead to a strategy in which you always bet that you are not a Boltzmann brain. That is, if you all adopt this strategy  you will behave as if you are maximizing utility with respect to a set of credences which assign zero credence to being a Boltzmann brain, and equal credence over all persisting individuals. So in this more realistic case, it  is not true that you are rationally compelled to adopt a strategy aiming for the greatest possible winnings summed over all subjectively identical centered worlds - there are clear practical reasons why you  might prefer to adopt a different strategy.

\subsection{Caring Measure \label{caring}}

A notable feature of the PSL cases above is that if we take it that the `rational' credences to assign are the credences $\{ p_i \}$ such that choosing a strategy which maximizes the quantity  $ W( \{ p_i \}) =  \sum_i p_i W_S(i)$ yields the highest total value of the goal quantity $G$, then once we have determined the goal  quantity $G$, the `rational' credences $\{ p_i \}$ are immediately fixed. If the goal quantity $G$ weights all the observers in the reference class equally it will  be rational to assign equal credence to all of them; if $G$ weights the observers proportional to mass it will  be rational to assign credences proportional to mass;  if $G$ excludes Boltzmann brains it will be rational to assign zero credence to all Boltzmann brains; and all of this is true  completely independently of any  empirical observations we might make if we actually perform the relevant experiment. 

This point can be generalized - in principle observers in this scenario could adopt any goal quantity  $G = \sum_i c_i W_S(i)$, where $c_i$ is an arbitrary set of  weights for the winnings. In fact, it can be shown that assigning negative weights $c_i$ leads to a scenario in which a `Dutch book' can be made\cite{schupbach}, which in the self-locating case means that agents seeking to maximize this quantity will accept bets which are guaranteed to result in all of them losing money. So perhaps we can make the case that  rational observers must not choose negative weights - but it would seem that any non-negative real number weights are permissible. And then  it follows immediately that if we say the rational credences to assign are the values $\{ p_i \}$ such that a strategy $S$ chosen so as to maximize $W( \{ p_i \}) = \sum_i p_i W_S(i)$ will yield the highest total value of  $G $, the rational credences will  always be given by $p_i =  c_i  \ / \sum_i c_i $.  

That is, once we have chosen a goal $G$ in a self-locating betting scenario, there is no point in doing empirical tests or any  kind of theoretical analysis to decide which credences lead to the best results with respect to that goal - the  credences follow immediately from the choice of goal.  So  if the `rationally compelling' credences in a PSL scenario are the ones which are optimal to achieve our practical goals, then  those credences are nothing more or less than a direct encoding of those practical goals. Thus in a sense,  the practical function of PSL credences in decision-making scenarios is to act as something like a  `caring measure,'  as proposed by Greaves in the context of the Everett interpretation\cite{Greaves}: from a practical point of view, PSL credences simply describe the extent to which we value winnings accrued by various observers in the reference class.

But if  PSL credences should really be understood as something like a caring measure, this undermines the idea that pragmatic rationality dictates how we ought to set them. For nobody is rationally obliged to care in a particular way, or at all, about the members of a certain class of subjectively identical observers; and thus if PSL credences are in practical terms just an encoding of our goals, it can't be the case that pragmatic rationality prescribes some specific way we ought to set them.    Any caring measure is `rational,' provided that it is probabilistically consistent: as Price puts it, `\emph{Rationality may dictate choice in the light of preference, but it doesn’t dictate preference itself}'\cite{price2008decisions}. So it appears that there is simply no space for pragmatic rationality to play any role in constraining our PSL credences, because as soon as we have chosen our goals, this immediately fixes how we ought to act.

Note that this criticism applies regardless of whether the observers in the reference class are    distinct observers, or different time-slices of the same observer - for in the latter case we are still free to make various different choices about how to prioritize gains accrued by different temporal parts of ourselves. This can be regarded as an instance of  Hedden's notion of `Time Slice Rationality,' based on the idea that `\emph{determining what an agent ought to believe does not require first figuring out the correct theory of personal identity over time}'\cite{Hedden2015-HEDTR}.  That is, the way we assign credences over a reference class surely should  not depend on whether or not the observers in the reference class are described as identical, so if the credences can be understood as a `caring measure' in a case which is naturally described as involving completely distinct observers, they can also be interpreted as a caring measure in a case which is naturally described as involving temporal parts of the same observer. In particular, it is   not the case that an observer subject to self-locating uncertainty over different time-slices of herself is rationally compelled to maximize winnings over all the time-slices just because the situation happens to have been described as one of identity between time-slices, since after all she would not be so compelled if the situation were not described as one of identity. 

Now, there is one possible response the proponent of PSL credences might make at this stage. This would involve arguing that the aim of  science is not to produce a description of reality which is true in an absolute sense, but rather to produce a description of reality which is convenient for some set of observers. Then it could be argued that the role of PSL credences in scientific practice is not to encode physical content, but rather to encode a stipulation about which set of observers the scientific reasoning in question is intended to work for - hence, it plays the role of a `caring measure' by dictating the set of observers that the reasoning is designed for.   Now, presumably not everyone would agree that this is really the aim of science, but in any case, if this is what proponents of PSL credences intend, then  conclusions drawn using PSL credences  really ought to be indexed to the practical context for which they are intended - so for example, rather than offering as a conclusion that `you should believe  you are in a simulation,' proponents of the simulation hypothesis ought to say `if you care equally about all subjectively identical copies of yourself, then you should assign a high self-locating credence to being in a simulation.' But such conclusions are  seldom presented in this way: they are usually presented as if they are rationally compelling in an absolute sense. So if PSL  credences are really to be thought of as simply encoding  practical priorities, a number of conclusions drawn using such credences should be moderated in very significant ways.

 \subsection{NSL and SSL vs PSL}

At this juncture, one might worry that the argument given above might also end up applying to NSL credences -  if that were the case then we would surely have to conclude that something has gone wrong somewhere! But as a matter of fact, the same argument does not carry over to the NSL cases, for  in NSL and SSL  cases, just fixing the goal of our betting procedure  does not already determine which credences will best achieve that goal. For example,  in the probabilistic case described in section \ref{betting}, if we say  that the rational credences to assign are the values $\{ p_i \}$ such that a strategy $S$ which maximizes $W( \{ p_i \}) = \sum_i p_i W_S(i)$  will yield the highest value for the goal quantity $G = \sum_j \sum_i W_S(i) \delta(i, O(j))$ summed over a large number of experiments, we can't immediately infer what those values $\{ p_i \}$ are directly from the formula for $G$: clearly  we will still have to perform empirical tests or theoretical investigations to determine which credences will in fact lead to the greatest winnings. 

One might  worry that things only come out this way because we are implicitly assuming that there is only one possible goal in an NSL case, i.e. to maximize the long-term sum of the winnings, whereas in the PSL case I have urged that there are many different reasonable goals. But in fact, even if we allow different possible goals in the NSL case, the same observation still applies. For example,   rather than just maximizing the total cumulative winnings in an NSL case, we could discount  winnings further into the future using a risk premium, so the goal quantity might be something like $G = \sum_j \sum_i W_S(i) \delta(i, O(j)) e^{-|c|j}$. But nonetheless, once we have chosen such a goal, we still typically need to appeal to experiment or theory to determine which credences do in fact yield the highest value for the quantity $G$ - it's not possible to determine those credences directly from the formula for $G$ given above. 

What, then, is the relevant difference between the NSL and the PSL cases? Well, the key point is that in a NSL scenario, we can identify observers from a third-person point of view  without saying which outcome they observe. For example, we can choose a description according to which there is just one observer who persists through the whole experiment and try to maximize her total winnings, or we can single out a particular temporal part of that observer who performs the sixth experiment in the sequence and try to maximize her winnings, and so on. The point is that after we have identified the observer(s) whose winnings we are trying to maximize, there is a further empirical question about what outcome(s) that observer does in fact see, meaning that we can  assign non-trivial credences to various possible outcomes  and then  determine either theoretically or empirically how close those credences are to the actual values or relative frequencies. This explains why pragmatic rationality can place meaningful constraints giving rise to nontrivial credences even after  our pragmatic goals have been fixed. 

By contrast, as discussed in section \ref{indexical},  in a genuine PSL case we can only identify observers by saying what centered world they are in, or by indexical self-reference.  Beginning with the first horn of this dilemma, in a PSL case the relevant `outcome' to which we are assigning credences is simply the centered world in which one is located, so if we identify an observer by saying what centered world she is in, then we have already determined her `outcome' and thus there is  no remaining empirical fact over which any nontrivial credences could be defined.  Meanwhile, taking the second horn of the dilemma, if we identify an observer indexically from a first person point of view, then there is indeed a further (self-locating) fact about what that observer will observe in a single experiment, but   then there is no possible pragmatic justification for  any non-trivial credences. One might initially imagine that we could repeat the experiment  to see which credences produce higher winnings for this indexically individuated observer over time - for example,  Bostrom suggests  `\emph{if we imagine the experiment repeated many times, the only way a given participant could avoid having a negative expected outcome when betting repeatedly against a shrewd outsider would be by setting her odds in accordance with SSA}'\cite{Bostrom2002-BOSABO-2}. But while this approach would work for NSL or SSL credences, it is impossible for PSL credences, because the definition of PSL credences requires that observers can only be identified either indexically or by saying what centered world they are in; whereas if it is possible to track `the same' observer across several experiments, then it would be possible to identify an observer without saying what centered world they are in during the current experiment, since we could simply point to the observer who obtained certain results in previous experiments and ask what centered world that observer is in now.   So in a genuine case of PSL uncertainty, indexically individuated observers cannot be identified across experiments, and thus each indexically individuated person only ever sees one outcome, so the empirical facts cannot possibly favour any credence other than   $1$ (for  the actual outcome) or $0$ (for all other outcomes). Thus either way, pragmatic constraints cannot yield nontrivial credences, and thus the only possible role for such credences is to encode something like a caring measure.
  
\section{The Principle of Indifference \label{analogies}}

If the claim that certain assignations of PSL credences are `rationally compelling'  cannot be understood in terms of pragmatic rationality, perhaps it is instead referring to epistemic rationality. But what could render certain assignations of such credences rationally compelling from the epistemic point of view?

Perhaps the most common way of defining epistemic rationality is in terms of aiming towards truth: `\emph{An epistemically rational agent must strive to hold a system of full beliefs that
strikes the best attainable overall balance between the epistemic good of fully
believing truths and the epistemic evil of fully believing falsehoods}'\cite{Joyce1998-JOYANV}. But here we encounter an immediate problem for the idea that epistemic rationality prescribes certain ways of assigning one's self-locating beliefs. For we already saw in section \ref{caring} that it is impossible to demonstrate empirically that certain PSL credence assignations are more pragmatically successful than others, and this also applies to demonstrating that certain PSL credence assignments are more likely to produce true beliefs than others: before we can say that assigning PSL credences in a certain way is a good way of coming to believe true things, we must specify for whom it is a good way of believing true things, and making that specification completely fixes which credences will best result in our chosen observers  believing true things. Thus much as in the pragmatic case, there appears to be little room for epistemic rationality to play any meaningful role in constraining credences once we have decided on an epistemic goal\footnote{One possible point of difference in the epistemic case is that one might think the initial choice of epistemic goal is more constrained than in the pragmatic case. For example, even if there is no pragmatic obligation to care in a certain way about Boltzmann brains, perhaps it can be argued that there is an epistemic obligation to take Boltzmann brains into account when considering what counts as a good way of arriving at true beliefs. But nonetheless, the same point applies. Even if epistemic considerations do place constraints on our epistemic goal in a PSL case, nonetheless once that goal has been chosen it entirely fixes the PSL credences, leaving no room for further empirical enquiry to determine the right credences. Thus it remains true that we cannot show empirically that some possible way of arriving at PSL beliefs is better for the purposes of believing truths. That leaves us with the possibility that there is some a priori principle of epistemic rationality that determines the correct epistemic goal and hence also the correct assignation of credences, which I discuss below. Thanks to an anonymous reviewer for raising this possibility.}.    

So instead of arguing that certain PSL credences are rational in virtue of leading to true beliefs, proponents of PSL credences typically argue that the  correct PSL credences are   determined   according to  certain principles. For example, such arguments often employ something like Elga's  Principle of Indifference\cite{Elga2004-ELGDDE} (PSL-POI) which says  that `\emph{similar centered worlds deserve equal credence}' (here the   term `similar' refers to centered worlds which belong to the same possible world and which are subjectively identical).  Since we cannot hope to demonstrate empirically that  the PSL-POI is a good way of coming to believe true things, the claim that we are rationally required to set our credences according to such a principle must presumably be interpreted as asserting that the PSL-POI  is something like an a priori principle of epistemic rationality.  But is it?   To answer this question, it will be informative to take a brief detour  to consider the status of a similar principle often employed in NSL cases.

\subsection{The Non-Self-Locating Principle of Indifference \label{NSLPOI}}

In scenarios involving non-self-locating uncertainty, it is common to employ a principle which is  sometimes referred to as `the principle of indifference,' (NSL-POI) or else `the principle of (in)sufficient reason,' which mandates that in the absence of any relevant evidence distinguishing between various mutually exclusive possible outcomes, we should distribute our credences equally between these outcomes\cite{Eva2019-EVAPOI}. One might be tempted to think that  the NSL-POI is  an a priori principle of epistemic rationality, in which case it would make sense to think its PSL analogue is also an priori principle of epistemic rationality. And indeed, there are various theoretical justifications one might offer for such an a priori principle  - for example, it can be shown that the NSL-POI is a special case of Jaynes' entropy principle\cite{PhysRev.106.620}, which may be interpreted as showing that it mimimizes bias, or that it is the minimally committed option\cite{schupbach}. 

However,   the idea that the NSL-POI is an a priori principle of epistemic rationality is undermined by Bertrand's paradox\cite{Shackel2007-SHABPA,schupbach}, which refers to the fact that there are generally different ways of dividing an outcome space up into individual `outcomes,' and applying the principle of indifference to different divisions will result in different probability assignations. A classic example is   Buffon's needle experiment, in which a needle is to be dropped onto the floor, and the task is to calculate the probability that the needle crosses the cracks between floorboards. One way to apply the principle of indifference here  would involve dividing the outcome space up with respect to the angle that the needle makes with the vertical axis; another possibility would involve dividing the outcome space up with respect to the vertical distance between the top and the bottom of the needle. And these two choices will result in different predictions for the probability of crossing the cracks, so we can't solve the problem using the  NSL-POI  without first making a choice about how to partition the outcome space\cite{VanFraassen2012-VANETR-2}.

Now, Bertrand's paradox is sometimes thought to apply only when the set of possible outcomes is continuous, so one might still try to argue that the  NSL-POI  is an a priori principle of epistemic rationality when the number of outcomes is finite. However, there is a sense in which the set of possible outcomes is continuous in all realistic scenarios, since  there will always be an (effectively) continuous range of possible final states for any real physical system. In some cases, such as rolling a die or flipping a coin, there is a particularly obvious way of dividing these final states up into discrete outcomes, but the mere fact that such a description exists does not guarantee that the right way to apply the  NSL-POI  is to assign equal credences to the outcomes thus described: for example, when rolling a die, I can choose to characterize the outcomes as `1' and `not 1,'  but this does not entail that we should assign probability 50\% to the outcome `1.'  So  in real physical situations we cannot simply take for granted that the outcomes as they are initially described are the right partition to use in applying the  NSL-POI: we must pay attention to  the  details of the actual physical situation in order to decide if the way of partitioning the outcome space provided in the problem description is  physically plausible. Thus Bertrand's paradox is  relevant even in cases which are initially described as if they have a finite set of discrete outcomes, because these may not always be the right set of outcomes to which to apply the principle of indifference.

So what is it exactly that makes for a good choice of partition in cases of NSL uncertainty? Well, in realistic examples it usually transpires that the partition which makes the best predictions is one which  reflects relevant features  of the process which determines the outcome - in particular, symmetries of that process\cite{Jaynes1973TheWP,van1989laws}. In the case of Buffon's needle, most experimenters will  drop the needle in a way which  is blind to rotation angle, since experimenters are typically not aiming at the cracks; and thus the right way to assign credences is often to use a distribution which is invariant under rotations, which amounts to applying the principle of indifference to a partition with all outcomes spanned by equal rotation angle\cite{van1989laws}. We  can verify empirically that this distribution matches the observed results for a needle dropped blindly. On the other hand, if we design the experiment differently by having experimenters deliberately aim at the cracks, then we should instead use a distribution which is not invariant under rotations, which will amount to applying the principle of indifference to a different partition.  

The key point is that the partition which will yield the most successful results in an application of the NSL-POI is not knowable a priori just from an abstract description of the outcome space - the success of the method depends on how closely the choice of partition reflects the features of the process by which the outcome is selected. Now,  by definition one is supposed to apply the POI to one's total evidence, so of course if you already know about the symmetries of the process selecting the outcome, you will only be considering partitions which reflect the constraint of indifference with respect to that evidence. But we often find ourselves applying the principle of indifference in cases where we do not know all the details of the process producing the outcome, and in such circumstances we can think of choosing a partition as an attempt to guess features of the process which we do not currently know. For example, in the Buffon's needle case, if you do not know any details about how the needle is being dropped, choosing a distribution which is invariant under rotations amounts to guessing that the needle is indeed being dropped blindly. 

Why then is the NSL-POI often successful in such cases? Well, in the kinds of sceanrios encountered in everyday life, it often turns out that the most intuitively natural partition of the outcome space   is indeed the one which corresponds to symmetries in the probabilistic process producing the outcomes.  Indeed, it's likely that our intuitions about the naturalness of  partitions have developed in such a way as to track the symmetries of common probabilistic processes occurring in our everyday lives. So in a case where you don't know all the details of the process producing the outcome, it's  often a reasonable first guess to simply apply the NSL-POI using  the partition that feels most intuitively natural, in the hope that this will turn out to match up to the symmetries of the relevant probabilistic process.  However, the fact that this methodology is often successful  does not mean that it is an priori principle of epistemic rationality: the  NSL-POI is a  rule of thumb which allows us to sensibly assign probabilities in cases when we have only limited information about the   underlying symmetries of the process producing the outcome, but it should subsequently be subjected  to actual experimental investigations in which we empirically establish the actual probability distribution and/or properly determine the nature of the process which produces the outcome. As  van Fraassen puts it, `\emph{This method always rests on assumptions which may or may not fit the physical situation. Hence it cannot lead to a priori predictions. Success, when achieved, must be attributed to the good fortune that nature fits and continues to fit the general model with which the solution begins}'\cite{van1989laws}.  

\subsection{The PSL Principle of Indifference \label{PSLPOI}}
 
These observations give us reason to question the status of the PSL-POI - for if the NSL-POI is not an a priori principle of epistemic rationality, why would that be different in PSL cases? In particular, one might worry that concerns along the lines of  Bertrand's paradox would apply in the PSL case as well.  Elga's formultion of the PSL-POI assumes we should use a partition of the outcome space in which each conscious observer corresponds to one  outcome, but although such a partition does seem natural,  there are certainly other possible partitions - for example, in the variant mass case discussed in section \ref{betting}, one could  partition the outcome space so as to have equal mass per outcome, leading to a probability distribution which assigns higher probability to observers with larger mass, rather than equal probability to all observers. 

Now,  Builes defends his principle of Center Indifference (a variant on Elga's principle of indifference) on the grounds that, unlike the NSL-POI case, it comes with a partition already specified: (NSL-POI)  `\emph{doesn’t specify a unique way one should partition the space of possibilities that one is indifferent over, but Center Indifference specifies that one should be indifferent between maximally specific similar centered worlds}'\cite{BuilesForthcoming-BUICIA}.  However,  the fact that Center Indifference has been formulated in this way  does not in itself guarantee that the specified partition is always right. After all, there are  many ways in which we could strengthen the NSL-POI to give a unique way of partitioning the space of possibilities in certain kinds of cases - for example, we might adopt a `Die Principle' stipulating that in  any case of uncertainty involving dice, one should always choose a partition in which each side of the die corresponds to a single outcome. But we don't typically deal with Bertrand's paradox by adopting such strengthened principles, because we recognise that in  the actual world, the correct choice of partition is not something which can be known a priori -   it must be determined empirically with reference to the real physical process producing the outcome. It would be a mistake to adopt the Die Principle as an a priori principle of epistemic rationality, because sometimes dice are weighted. So it is no virtue of PSL-POI or Center Indifference that they specify a way of selecting a partition, unless we can  give some reason to think  this particular partition  is in fact uniquely rationally compelling.

Given   the similarities between the PSL and NSL principles of indifference, one might naturally think that the right way to choose a partition in the PSL case would be similar to the   NSL case, meaning that it would involve trying to match  the symmetries and other features  of the process which produces the outcome. But here we arrive at an important disanalogy. For in a PSL case  the `outcome' - i.e. which centered world an indexically individuated observer turns out to be located in - is not produced by any physical process, since the observer is not literally dropped into one location rather than another.  Therefore we cannot determine the right partition by appealing to features of the  process producing the outcome,  since there is no such process. And therefore one main justification for using the principle of indifference  in an NSL case is  absent in the PSL case - it doesn't make sense to appeal to a rule of thumb which works by  helping us guess the underlying symmetries of the process which produces the outcome if there is no such process in the first place! 

Of course,  there will typically  be some symmetries present in the outcome space for a PSL scenario, or in the process which produces the relevant set of observers in toto. And indeed, there have been attempts to use such symmetries to justify either the PSL-POI, or some  other way of assigning self-locating credences.  For example,    some Everettians have argued that the assumed preference for applying the  principle of indifference to a partition with one consciousness per outcome can be overridden by knowledge of symmetries.  In particular, Sebens and Carroll argue for the Epistemic Separability Principle:  `\emph{ESP: The credence one should assign to being any one of several observers having identical experiences is independent of the state of the environment}'\cite{10.1093/bjps/axw004}, which amounts to requiring that our credences should be invariant under transformations affecting only the environment, which are taken to be symmetry transformations. Similarly, Vaidman and McQueen adopt a principle requiring that when an experiment respects a symmetry, it will lead to a symmetry between descendants corresponding to the measurement outcomes\cite{pittphilsci15195}.

However, although it is true that in the Everettian scenario there are certain symmetries present in the process which creates the set of post-measurement branches and observer  as a whole,  a process which produces the set of centered worlds in totality is importantly different from a process in which a specific observer is placed into one particular centered world rather than another, and clearly there is no process of the latter kind in the standard Everettian picture, nor in any other PSL case. And note that in the NSL case, the mere existence of symmetries in the general vicinity of the relevant scenario is not enough to tell us what probability distribution we ought to adopt. For example, in the outcome space for the Buffon's needle case we can identify various possible symmetries of the outcome space, including a possible rotational symmetry, which is encoded in the `equal angle' partition, and a possible translational symmetry in the direction orthogonal to the floorboard cracks, which is encoded in both the `equal angle' and the `equal distance' partition. But we cannot determine a priori that the appropriate probability distribution should be invariant under one or both of these symmetries. To establish that, we have to consult the details of the actual process by which the outcome is generated in order to determine which symmetries are  relevant to the way in which the outcome is actually determined - and if we change that process by dropping the needle in a different way,  the appropriate probability distribution will change, even though the outcome space and the rest of the experimental setup remains the same.  So it is the symmetries of the process producing the outcome, rather than just general featuers of the setup, which are important here.  

 Yet in a PSL scenario  there cannot be any such process; and in the absence of such a process, there is no possible way of demonstrating any direct link between  general symmetries of the experimental setup and the credences  one should assign over finding oneself in various locations, since none of these symmetries play any role in determining   which location one finds oneself in.  Symmetry-based arguments such as those of refs \cite{10.1093/bjps/axw004, pittphilsci15195} may initially  look compelling, but this is at least to some extent because they appeal to intuitions developed in NSL cases, in which it is  a reasonable first guess to hypothesize that the process producing the outcomes in a given scenario may be invariant under `natural' symmetries of the experimental setup or  outcome space. But in the NSL case this is simply a rule of thumb which stands in for actual knowledge of the process producing the outcome, so its use in NSL cases offers no justification for using the same kind rule in PSL cases where there cannot be any  such process to ground the reasoning. 
 
\subsection{Factors Specific to the Centered Case}

It appears that the kinds of factors which  determine the appropriate choice of partition for typical applications of the NSL-POI are not present in putative applications of the PSL-POI. So if there is nonetheless a rationally compelling way to apply the POI in PSL cases, it is most likely  determined by factors which are specific to PSL cases. What could those factors be?

 Perhaps the most obvious point of difference between the NSL and PSL cases is that in the PSL case outcomes are  attached to centered worlds defined by distinct consciousnesses, rather than to subdivisions of a set of physical states. So one may be tempted to argue that there is something about the nature of consciousness itself which means that we are rationally compelled to apply the principle of indifference using a partition with one outcome per consciousness.   But it's unclear that this is always the right result - for example,   we saw in section \ref{PSLPOI}  that in the Everettian case it is often argued that a naive application of the principle of indifference using  `one consciousness per outcome' is not  correct.  So if we are willing to entertain these kinds of arguments,  we are by implication accepting that there is no a priori principle of rationality which mandates that we must always  assign equal credence to every  consciousness. This suggests that we must  determine the right way of assigning credences by appeal to the features of the actual physical situation - and yet, as we saw in section \ref{PSLPOI}, the kind of features which determine the credences in the NSL case are absent in the PSL case, and it's unclear that there is any suitable replacement for them. 
 
 Moreover, even if we are willing to disregard the Everett interpretation and other related  cases in order to insist that the partition with one outcome per consciousness is always correct, we run into difficulties when we try to justify that choice, because treating consciousness itself as a fundamental constraint on the way in which we should partition a physical outcome space looks uncomfortably close to letting dualism in through the back door. From a physicalist point of view it  is natural to think that we should be able to define rational constraints on beliefs without making essential reference to a vague non-physical notion like `consciousness,' and yet there is surely  no precise way to define the `one outcome per consciousness' partition without invoking  consciousness. Of course,  since beliefs themselves are  not a fundamental entity, we should not necessarily expect that normative constraints on beliefs will  invoke only fundamental physical objects; but nonetheless it seems natural to think that normative constraints on beliefs should at least be  expressible  in physical terms, even if there is a more succinct way of expressing them using non-physical notions like `consciousness', whereas it seems very unclear that the norm demanding a partition with one outcome per consciousness could ever be formulated in purely physical terms. Thus it remains unclear that such a partition can be justified without  assigning a privileged physical role to consciousness. 
 
  One possible justification is suggested by Builes, who argues for  Center Indifference on the grounds  that `\emph{the usual reasons for why one might favor one possibility over another don’t seem to be present in Center Indifference}'\cite{BuilesForthcoming-BUICIA}. Now, an immediate problem with this is that Builes appears to be presupposing a choice of partition rather than offering any argument for it - for if we were to choose a partition which makes  finer subdivisions of the centered worlds, the elements of that partition would presumably  still have the property that there is no reason to favor any of them over any other, so Builes' approach makes sense only if we have already decided that a partition with one outcome per consciousness  is the only option. 
 
 But in addition, is the absence of any possible reasons really favourable to Center Indifference? Builes focuses here on reasons pertaining to theoretical virtues, noting that the PSL hypothesis that I am in one centered world cannot be simpler or more explanatory than the hypothesis that I am in another centered world. But this point can be taken further - we saw in section \ref{PSLPOI} that in PSL cases there also cannot be any reasons arising from the nature of the process determining the outcome which would favour one possibility over another.  So in the PSL scenario, it does not just happen to be the case   that there are no reasons favoring either of the outcomes - there is simply no kind of reason which could   possibly favour one outcome over another, and thus  our credences in this scenario are completely unconstrained by any `reasons.' 
 
 Note that this is markedly different from NSL cases. In NSL scenarios, our applications of the principle of indifference do not typically have the feature  that there are  no possible kinds of reasons which could ever lead to one outcome being favored over the other - rather it just happens to be the case that in some particular instance the `reasons' present favor all of the outcomes equally. This is important, because it means that in general, if we consider some alternative partition of the outcome space then it will typically no longer be the case that the reasons present favor all of the outcomes equally, and thus there is an objectively correct way to decide which partition is the one to which we should apply the NSL-POI - i.e. the one for which the reasons 
do  favor all of the outcomes equally.  Whereas in the PSL case, no matter how we partition the outcomes there will never be any reasons favoring one outcome over another, and thus there is no fact of the matter about which partition is the one to which we should apply the PSL-POI, since they are all equally good in this regard. So it is quite unclear   that we ought to respond to a scenario in which there could not possibly be any reasons favouring one choice over another in exactly the same way as we respond to a scenario in which the reasons present happen to favour all outcomes equally. One might think that in the former case   the right response is to simply accept that there is no rationally compelling assignation of credences, precisely because there is no possible way in which any `reasons' could ever constrain such credences.  

Builes also offers a second argument: `\emph{Another way to support Center Indifference is by noting that violations of Center Indifference require a strange kind of forced epistemic disagreement. Suppose you deviated from Center Indifference in some way, say by being more confident in c1. Then, so long as you are self-aware, it will be implied by your evidence that you are more confident in c1. This implies that your evidential twin will also think that they are more likely to be located in c1}'\cite{BuilesForthcoming-BUICIA}. This argument draws on the natural intuition that agents with the same total information and the same epistemic standards should always end up assigning the same credences. However, this makes sense  only in scenarios where we have already accepted that there exists a uniquely rational way of assigning PSL credences. In that case,  if two agents assign contradictory credences, at least one of them must be  irrational. But if we are in a scenario where we have  accepted that there is no uniquely rational way of assigning credences - for example, in a scenario where we understand credences to play the role of something like a caring measure - then it wouldn't make sense to insist that similar agents are rationally compelled to end up with the same credences, for it has been admitted already that rational agents are free to assign credences in any way they like. In such a scenario,  disagreements between agents do not indicate irrationality - if agents may freely  choose how to assign these credences, then the fact that their choices do not agree does not imply that either of them is wrong. Thus this kind of argument does nothing to show that there exists a rationally compelling way of assigning credences in the first place, so it also does not prove that any particular assignation is rationally compelling.

\section{Analogical Arguments \label{analogies2}}

Because the  PSL-POI is similar to the NSL-POI,   arguments for the PSL-POI can be regarded as instances of a more general strategy in which it is argued that certain distributions of PSL credences are rationally compelling in virtue of an  analogy with structurally similar NSL or SSL cases. Moreover, the problems we have encountered in discussing the PSL-POI generalize to other such analogical arguments.  For in using analogies between two kinds of scenario to establish what is  `rational' in one of those  scenarios, it is important to first consider whether any possible disanalogies between the scenarios might undermine  the comparison, and   we have just seen that there is indeed a potentially fatal disanalogy between NSL/SSL and PSL scenarios: in the NSL/SSL case there is a physical process which produces the outcome over which we are assigning credences, so there are  facts about the process and its symmetries which ground certain rationally compelling assignations of credences over the outcomes, but in the PSL case there is  no such process, and thus  the kinds of  facts which often ground the rationally compelling credences in the NSL case are simply nonexistent in the PSL case.  This means that   the reasons we have for assigning certain credences in NSL cases will not in general be the same as the reasons we might have for assigning credences in PSL cases,  so  we should not assume  that credences from NSL cases will automatically transfer across to  PSL cases, even if they are structurally similar. 

 We can see an example of the analogical strategy in  Elga's argument for the PSL principle of indifference\cite{Elga2004-ELGDDE}. This argument uses  a chain of reasoning such that at each step of the chain  we are asked to agree that two scenarios are relevantly similar, so   the rational credences for one case can be inferred from the rational credences for the other case.  In particular, Elga considers scenario TOSS\&DUPLICATE in which an agent Al and a subjectively identical duplicate are put to sleep, then  a coin is flipped (with probability 10\% to land on Heads), and then both the agents are woken. Elga also considers another scenario, COMA, which is similar to TOSS\&DUPLICATE except that only one agent is woken and that person is given the information `\emph{either the coin landed on Heads and you are Al, or the coin landed on Tails and you are Dup.}' Elga argues that in COMA, the agent should assign probability 10\% to the coin landing on Heads.  Elga contends that this means that in TOSS\&DUPLICATE the agent should assign credence 10\% to $p(HEADS | HeadsAl \vee TailsDup)$, and this can then be used to calcualte the credence the agent should assign to $Al$ by manipulating the standard conditional probability formula. 
 
 But we should be careful here, because the additional information provided in the COMA case shifts us from a PSL case to a SSL case. In scenario  TOSS\&DUPLICATE there are two possible worlds in which the agent could be located: $W_H$ and $W_T$ in which the coin lands Heads and Tails respectively. And there are four centered worlds in which the agent could be located: $(W_H, C_A)$,  $(W_H, C_D)$,  $(W_T, C_A)$,   $(W_T, C_D)$, with $C_A$ corresponding to Al and $C_D$ corresponding to Dup. Thus the agent has  a mix of NSL credences over the two possible worlds $W_H, W_D$, and then PSL credences distributed over the centers  $C_A$ and $C_D$ corresponding to Al and Dup respectively within those two worlds. But in the COMA case all the credences are SSL, since  the centered worlds  $(W_H, C_D)$,  $(W_T, C_A)$ are ruled out and thus now there are only  two possible centered worlds,  $(W_H, C_A)$,   $(W_T, C_D)$, which belong to different possible worlds. Thus given the important disanalogies between PSL and SSL cases, it should not be  assumed too readily that credences from COMA can be directly transferred over to TOSS\&DUPLICATE.

In particular, Elga's argument requires us to define the value $p(HEADS | HeadsAl \vee TailsDup)$ using the standard conditional probability formula applied to probabilities defined in the pure self-locating scenario TOSS\&DUPLICATE, but then also to assume that this value is the  same as the credence that an agent should assign to Heads in the superficially self-locating COMA scenario.  This assumption is motivated by the idea that a conditional probability of the form   $p(X|Y)$ represents  the credence you should have in $X$ if you come to know $Y$, with COMA being used as a concrete physical realization of a scenario in which the agent comes to know $(HeadsAl \vee TailsDup)$. But this way of thinking about conditional probability is only an interpretation, not a definition, and although it is a reasonable interpretation for most NSL cases, it should not necessarily be presupposed for PSL cases. For the actual mathematical definition of $p(HEADS | HeadsAl \vee TailsDup) $, as employed in Elga's argument, is  given by the conditional probability formula  $p(HEADS | HeadsAl \vee TailsDup) = \frac{p(HEADS \& (HeadsAl \vee TailsDup) )}{ p(HeadsAl \vee TailsDup)}$; and note that   if the thesis of this article is right, the assignation of  credences over Al and Dup  are not rationally constrained by anything other than probabilistic consistency. That means there cannot be some rationally compelling fixed value for the ratio $\frac{p(HEADS \& (HeadsAl \vee TailsDup) )}{ p(HeadsAl \vee TailsDup)}$, because if there were such a fixed value, that value would place constraints on the way in which we could assign credences to Al and Dup.  Yet there is  a rationally compelling way to assign a credence to HEADS if you come to know $HeadsAl \vee TailsDup$, since in that case you are in a SSL scenario, and  there are usually rationally compelling ways to assign credences in SSL cases. So it does not make sense in this kind of scenario to suppose that the value $p(HEADS | HeadsAl \vee TailsDup)$ calculated from the conditional probability formula is the same as  the   credence the agent should assign to HEADS if they come to know $HeadsAl \vee TailsDup$, since the latter has a unique value and the former cannot have a unique value. Thus the interpretation of conditional probabilities on which Elga relies simply breaks down in the move from the PSL case to the SSL case, and  therefore we are not obliged to transfer the credences over from one case to another as Elga's argument demands.

With that said, let me acknowledge that  there is a deflationary way of reading analogy-based arguments like Elga's on which the move from a SSL scenario to a PSL scenario is indeed reasonable. That is, we could  think of Elga's principle of indifference as simply aiming to characterize the way in which it is intuitively natural for agents like us to assign credences.  Then it may be argued that since the self-locating aspects of the scenario TOSS\&DUPLICATE are outside of our usual experience of probabilistic reasoning, the most  natural thing for us to do is to reason as we would in the most closely analogous  SSL case, which is probably something like COMA. So Elga's chain of reasoning may well be successful if the goal is just to arrive at a statement about the way in which it is intuitively natural for beings like us to assign credences. 

But the problem is that the principle of indifference is not usually understood as merely characterizing  natural intuitions; it is frequently invoked as a scientifically weighty principle from which significant  conclusions can be drawn.  Yet if it can only be understood as characterising  reasoning which feels natural to us,   then the   PSL credences it recommends are surely not   sufficiently objective to be used in these   scientific applications. So while the arguments of this section do not necessarily entail that we should refrain from employing the principle of indifference when we find ourselves in scenarios of PSL uncertainty, they do indicate that we should be  cautious about deriving any serious scientific conclusions from it. 

\subsection{Certainty \label{certainty}}

A particular subspecies of analogical arguments involves making comparisons to cases involving certainty.   For example, suppose that in Case A I know that none of the subjectively identical observers that I could possibly be are simulations, while in Case B I know that one of the subjectively identical observers that I could possibly be is a simulation, while the remaining 999   are not simulations. It seems natural to say that in Case A I am entitled to be certain that I am not a simulation;  but Case A and Case B are extremely similar, so surely if I am entitled to have certainty in Case A, I am entitled to have credence very close to 1 in Case B? We could then imagine moving through something like a sorites series to arrive at a more general argument for something like the principle of indifference\footnote{Thanks to Kelvin McQueen for suggesting this argument}. 

In response to this argument, note  that there are two importantly different ways of understanding the claim in case A that `I am certain that I am not a simulation.' It could be understood as a self-locating claim, of the form $P_1$: `I myself am one of the observers in my reference class who is not a simulation.' But it could also be understood as a non-self-locating claim, of the form $P_2$: `My experiences are not compatible with being a simulation.' Assuming we are able to provide a non-indexical characterisation of the nature of the relevant experiences, credences assigned to $P_2$ can be understood entirely from a third-person point of view - for example, we might arrive at them in an entirely impersonal way on the basis of hypotheses about what kinds of experiences are possible for simulations. 

And if we focus on proposition $P_2$ rather than $P_1$,    Cases A and B do not look so similar. For clearly in Case B I am obliged to assign credence 0 to proposition $P_2$; whereas in Case A there is some room for debate, but arguably in that case I should assign credence 1 to $P_2$. For any first-person evidence I might have which provides evidence for the proposition that `No observer subjectively identical to me is a simulation' must  also be part of the first-person evidence available to  all observers subjectively identical to me. So if my experiences are not incompatible with being a simulation, then it is possible for an observer to have this very evidence while being a simulation, and thus this evidence  cannot be  reliable evidence for the claim that no observer subjectively identical to me is a simulations. So plausibly the only way I can come to be   certain that no observer subjectively identical to me is a simulation is by coming to be certain that my experiences  are incompatible with being a simulation, i.e. by assigning credence 1 to $P_2$.

Thus there appears to be a discontinuous change in the credences assigned to the proposition $P_2$ between cases A and B, despite their apparent similarity. Moreover, this is true even if we increase the ratio of non-simulations to simulations in case B - the  credence we assign to $P_2$ will not approach 1 as this ratio approaches infinity. And therefore if `certainty' in case A is interpreted not in terms of the self-locating claim $P_1$ but in terms of the non-self-locating claim $P_2$, it follows immediately that even though I am entitled to be `certain' in case A, I am not obliged to be `close to certain' in case B.

So although it may seem counterintuitive to make such a strong distinction between cases  where all relevant observers have some property and cases where almost all relevant observers have some property,  this distinction is perfectly  reasonable once we recognise that `certainty' in case A need not be understood as  just a limiting case of PSL credence - it is arguably better analysed in terms of non-self-locating claims about the compatibilty of my experiences with being a simulation, and thus it does not imply anything about how we ought to distribute PSL credences in either case A or case B.

 \section{Scientific Applications \label{applications}}
 
Suppose we accept that there is no   rationally compelling way of assigning PSL credences. If this is true, it will have  consequences  for various common applications of   PSL credences, particularly  in scientific contexts.

The PSL principle of indifference and similar principles like Bostrom's Self-Sampling Assumption (SSA) are commonly invoked in various scientific debates.  Bostrom argues that we should see such principles as  `\emph{methodological prescriptions. They state how reasonable epistemic agents ought to assign credence in certain situations and how we should make certain kinds of probabilistic inferences}'\cite{Bostrom}. But if these methodological prescriptions are not rationally compelling, nor susceptible to empirical verification, what exactly are their credentials as methodological prescriptions? At one point Bostrom considers the possibility that SSA may not be a requirement of rationality, arguing that even so, `\emph{It suffices if many intelligent people do in fact - upon reflection - have subjective prior probability functions that satisfy SSA. If that much is acknowledged, it follows that investigating the consequences for important matters that flow from SSA can potentially be richly rewarding}'\cite{Bostrom}. But it seems possible that intelligent people may have these subjective probability functions only because of intuitions that have been inappropriately transferred  from SSL or NSL cases - so it may indeed be interesting to investigate the consequences of these probability functions, but we must be very careful about what exactly has been achieved in such an analysis. If the SSA, the PSL-POI and so on are not rationally compelling, and this is all just a matter of what `feels right,' we should be cautious about using this kind of reasoning to make strong claims about reality.

Of course, it should be emphasized that  many `self-locating credences' appearing in practical or scientific applications are in fact merely SSL credences, and thus the arguments of this article do not threaten such applications. For example, Bostrom describes a way of using self-locating credences to predict how fast cars will move in different lanes, based on treating yourself as a random sample from the set of all drivers on the motorway\cite{Bostrom}. This case is  an instance of SSL rather than PSL   uncertainty, because there is a causal history which results in you being in one position rather than another in the traffic jam, and thus different possible positions that you could have in the traffic jam correspond to centered worlds in different possible worlds, rather than different centered worlds within the same possible world. So the arguments I have made in this article don't undermine this kind of reasoning. 

However, there are certain scientific contexts in which it is common to invoke probabilities which appear to be genuine PSL credences, and thus the arguments of this article do pose a threat to those applications. I will now discuss several such applications.   Evidently a possible strategy one might adopt in response to my concerns would be to try to show that the credences involved in these cases can in fact be understood as  SSL rather than PSL in nature - and indeed I think this would be an interesting route to explore, but I do not have space do so here, so in what follows I will simply take for granted that the credences in these cases are in fact PSL credences.

\subsection{Personal Circumstances \label{Boltzmann}}

One common application of PSL credences in science involves using such credences to draw conclusions about your personal circumstances, as in claims such as `you are very likely to be a simulation'\cite{Fallis2023-FALSAS-2}, `you are very likely to be a Boltzmann brain'\cite{BuilesForthcoming-BUICIA}, or, in the Doomsday argument\cite{Bostrom}, `you are likely to have been born at around the midpoint of the birth order of all humans who will ever exist.'

For example, arguments for the simulation hypothesis\cite{Fallis2023-FALSAS-2,2ac867a4-6887-3270-8211-53d9c0e24445} typically start off by asserting that we have good reasons to believe that the world contains many more simulations than actual people, and therefore that large numbers of these simulations may be having experiences subjectively identical to yours. Then  the PSL-POI is invoked to argue that you should believe you are most likely in a simulation. However,  the self-locating credences here appear to be PSL credences, since there is no physical process by which you are dropped either into a real person or a subjectively identical simulation. So if  there is never any rationally compelling way to assign PSL credences, then there cannot be a rationally compelling way to assign a credence to being a simulation in this scenario. Thus although it would be permissible to assign high credence to being a simulation, it is equally permissible to assign high credence to not being a simulation, and therefore the simulation argument by itself does not establish very much\footnote{One might seek to avoid this problem by considering a reference class of simulated and non-simulated observers who are not subjectively identical, in which case we are not in a PSL scenario. But then the ratio of simulations to non-simulations is significantly less relevant to our assessment of our situation, since we can alternatively base our credences on the  compatibility of our own experiences with being a simulation, without reference to  other observers. Thus the simulation argument still runs into problems in this case, since   it is no longer clear that the high ratio of simulations to non-simulations in and of itself  entails that we must assign high credence to being a simulation.}.

Much the same applies to the Boltzmann brain case. If you believe that the world likely contains many more Boltzmann brains than persisting human individuals, then  a straightforward application of the PSL-POI suggests you  are most likely a Boltzmann brain. But again, the credences here appear to be PSL credences, since there is no physical process by which you are dropped into a real person or a Boltzmann brain. So while it would be reasonable to assign high  credence to being a Boltzmann brain, it would also be reasonable to assign low  credence to being a Boltzmann brain - and indeed in section \ref{betting} we saw that from a pragmatic point of view this seems like quite a reasonable thing to do. Thus again, we are not rationally compelled to believe that we are probably Boltzmann brains, so this  argument by itself does not establish very much.

\subsection{Empirical Confirmation \label{confirmation}}

The second kind of application involves using pure self-locating credences to perform empirical confirmation by means of Bayesian updating. This occurs in some multiverse scenarios\cite{Azhar2015-AZHTTI}, but is perhaps most prominent in the Everett interpretation. In that context  it is often  argued that mod-squared amplitudes should be interpreted as (pure) self-locating credences for finding oneself in one branch of the wavefunction rather than another, and also  that inhabitants of an Everettian world can perform Bayesian updating based on observed measurement outcomes by using these PSL  credences in exactly the same way as we normally use non-self-locating probabilities\cite{10.1093/bjps/axw004,pittphilsci15195}. For example, this means that if I am considering two versions of an Everett-style  theory which assign different mod-squared amplitudes to a certain measurement outcome, and then I do in fact see that outcome, I ought to update my credences to assign higher probability to the version of the theory which has a larger mod-squared amplitude for that outcome, following  the usual Bayesian updating formula. It should be noted that since empirical confirmation involves deliberating over multiple possible world, in this scenario we are necessarily dealing with the more realistic case alluded to in section \ref{superficial}, rather than the simpler idealized scenario I have been discussing throughout this paper.

 Now, one reason to think there may be something wrong with this approach to empirical confirmation follows from a view that Titelbaum calls the Relevance-Limiting Thesis (RLT)\cite{2ac867a4-6887-3270-8211-53d9c0e24445,Bradley2007-BRABAS}, which claims that learning a piece of  self-locating information should never cause us to update our non-self-locating credences. Evidently the RLT entails that self-locating information cannot be used in empirical confirmation, since empirical confirmation  is about updating non-self-locating credences assigned to  various hypotheses.  There are good reasons to think the RLT may be true, because  approaches to belief-updating which do not uphold the RLT typically lead to extremely counterintuitive results in cases where the number of subjectively identical observers in a given world can increase over time\cite{Meacham2008-MEASBA, Bostrom2007-BOSSBA-3,Fallis2023-FALSAS-2,AdlamEverett}. But nonetheless, a number of philosophers (including Titelbaum himself\cite{2ac867a4-6887-3270-8211-53d9c0e24445,Bradley2007-BRABAS}) believe that  the RLT is false. 

I think these disagreements over the RLT may arise from a failure to distinguish properly between  PSL and SSL scenarios, for there is a very straightforward intuitive argument which suggests that the RLT is true for PSL information, but not for SSL information. If you learn pure self-locating information, then learning that information only tells you which centered world you are in from a set of centered worlds all belonging to the same possible world, so it cannot tell you anything new about which possible world you are in, i.e. it cannot change your non-self-locating credences.  Whereas if the information you learn is only superficially self-locating, then when you learn which centered world you are in you also learn which possible world you are in, so clearly you do have reason to update your non-self-locating beliefs.    

And indeed, if we examine putative counterexamples to the RLT,  at least the most obvious kinds of cases turn out to concern SSL credences rather than PSL credences. For example, in the case considered in section \ref{superficial} about knowing the time upon waking, the unqualified RLT would seem to suggest that I shouldn't update any non-self-locating beliefs when I check the time and see that it is seven o'clock, but that is surely wrong - on seeing that it is seven o'clock I learn how long a certain human being slept on a given occasion, and that  can potentially lead me to update various  non-self-locating beliefs about the state of health and sleep hygiene of that human being (who happens to be me, but no part of the belief-updating I am doing rests upon this fact). However, the information I learn when I check the time is not PSL information but only SSL information, since it tells me whether I am in a possible world where a certain human being slept for six hours or an alternative possible world where that human slept for seven hours. So this counterexample supports the view that the RLT is false for SSL information, but this is perfectly compatible with the hypothesis that the RLT as it pertains to PSL information is correct.

With that said, the argument given above is of course an oversimplification, for the point of empirical confirmation is that it pertains to a  scenario where you do not know exactly  which possible world you are in, so any instance of empirical confirmation which involves  deliberating over a range of centered worlds must actually pertain to the more realistic case where there  exist (subjective) duplicates of those centered worlds in various different possible worlds. So the real question is, if there is a set $\{ P_1, P_2 ... P_N \}$ of possible worlds to which you assign non-zero credence, where each $P_{i}$ includes a set $\{ C^1_{P_i},  C^2_{P_i}... C^M_{P_i}  \}$ of   centered worlds that you could be located in, and you then learn  a piece of pure self-locating information $X$ which tells you that you are in the set of centered worlds $\{ C^X_{P_1}, C^X_{P_2} ... C^X_{P_N}\}$ but which does not give you any independent information about which possible world you are in, can this nonetheless cause you to update the credences you assign over the possible worlds in  $\{ P_1, P_2 ... P_N \}$? 

 The RLT suggests that it cannot, but here is one approach one might take to argue that the RLT is wrong. Imagine that the worlds $P_1$ and $P_2$ have laws or symmetries which entail different assignations of PSL credences to their corresponding  centered worlds  $C^1_{P_1},C^1_{P_2}$, and suppose the PSL credence to find myself in $C^1_{P_1}$ mandated by the laws of world $P_1$ is higher than the PSL credence for $C^1_{P_2}$ mandated by the laws of $P_2$. Then suppose I  learn a piece of pure self-locating information which tells me that I am in a centered world in the set $\{C^1_{P_1},C^1_{P_2}\}$. Surely   in that case I ought to update my NSL credences to assign higher credence to $P_1$ and lower credence to $P_2$, thus changing my non-self-locating beliefs?  If this is right, the RLT seems to be false even for PSL information. 

    However, the thesis of this paper is that, given a set of centered worlds all belonging to the same possible world, there is no    rationally compelling way of assigning pure self-locating credences of those worlds. And if this is true, it follows that   laws or symmetries cannot entail anything about PSL credences, since otherwise there would sometimes be a rationally compelling way of assigning credences over centered worlds all belonging to a given possible world, based on features of that world's laws or symmetries. For example,  we examined the symmetry case in  section \ref{PSLPOI}, and concluded that the conditions for symmetries to mandate certain assignments of credences are not met in the PSL case, since in PSL cases symmetries do not ever play any role in determining the `outcome,' i.e. the centered world in which one finds oneself. If this is right, then  the symmetries that hold in  a given possible world cannot possibly entail any particular PSL credences that one ought to have in that world, and much the same goes for laws, and thus  the situation described above simply cannot ever occur.  This suggests that the RLT as it pertains to PSL information is indeed correct: although there might sometimes be  certain choices of PSL credences which feel intuitively natural given  a certain set of laws and symmetries,  if they are only intuitively natural as opposed to rationally compelling, then it would be a mistake to use them in Bayesian updating as if they are the same as ordinary non-self-locating probabilities, and thus learning PSL information can't cause us to change our non-self-locating credences in the way described above.

Additionally,   if we agree there is never a rationally compelling way to assign PSL credences, but we think there is sometimes a rationally compelling way to assign NSL credences, then in order to maintain the rationality of our NSL credences we should avoid allowing them to be `infected' by the subjectivity of our PSL credences, and thus we have good reason to adopt an approach to belief-updating which keeps NSL and PSL credences clearly distinct. That is,  we should probably adopt an approach to belief-updating  in which we `\emph{first assign (NSL) credences to possible worlds and then somehow distribute those credences over the centered worlds corresponding to the possible worlds}'\cite{AdlamEverett}, such as   the system proposed by Halpern and Tuttle\cite{10.1145/153724.153770} or Meacham's compartmentalized conditionalization\cite{Meacham2008-MEASBA}\footnote{However, note that it follows from the thesis of this article that there is no uniquely rational way of performing the second step distributing credences over centered worlds, so although either of these systems would be rationally permissible, neither is rationally required - we are always free to choose a different way of performing the second step.} And clearly any such approach to belief updating  will automatically uphold the RLT as it pertains to pure self-locating information.

Bradley\cite{Bradley2011-BRAHBM} makes a somewhat similar point, arguing that the RLT is true if it pertains to `Mutation - belief change in virtue of a change in the truth-value of the content of the belief' but false as it pertains to `Discovery -  belief change in virtue of the discovery of the truth of the content of the belief, where the truth-value did not change over the period of interest.' An example of Mutation is watching the hands of a clock move and changing one's beliefs about the time (because the statement `it is now twelve o'clock' changes in value) and an example of Discovery is being uncertain about the time and then looking at one's watch.  Applying the schema I have used here, we can see that examples of Mutation typically involve gaining PSL information, whereas examples of Discovery often seem to involve learning SSL information - for example, when you look at your watch, you do not just learn which centered world you are in, you learn that certain events (the event of you checking your watch, or other events going on around you as you check your watch) occur at twelve o'clock, so you learn that you are in a possible world in which those events occur at twelve o'clock rather than at some other time. Thus Bradley's way of distinguishing the cases in which the RLT is true from those in which it is false would likely agree with my approach in many instances. However, Bradley is particularly concerned with cases where the uncertainty is about one's temporal location, and his Discovery vs Mutation categorisation does not seem so straightforwardly applicable to other kinds of cases, such as being uncertain about which one of a set of subjectively identical clones one is at a certain fixed time.   So the difference between  PSL   and  SSL scenarios looks like a more  generalizable way to distinguish cases in which  the RLT is true.  

In summary,  if it is   accepted that there are no rationally compelling assignations of PSL credences, this suggests strongly that the RLT is correct as it pertains to PSL information. And if this is the case, it immedately follows that PSL credences should not be used to do empirical confirmation, either in the Everettian context or in any other context. Thus this line of argument provides further reason to be wary of the use of PSL credences in scientific applications.

\subsection{Anthropic Reasoning}

The third kind of application involves using PSL credences in anthropic explanations. For example, it has been proposed that we can explain the apparent fine-tuning of various fundamental parameters by first assuming we are in a certain kind of multiverse, and then arguing that in such a multiverse the appropriate self-locating credence to assign to  finding oneself in a universe with  fundamental parameters in the relevant range is   relatively high\cite{friedbook, Azhar2013-AZHPAT}.

Now, it is well known that this approach runs into problems if the multiverse in question is infinite, since in that case we must choose a measure  to determine the relevant self-locating credences, and there seems to be no rationally compelling choice of measure\cite{friedbook,Linde_2007,Bousso_2007}. However, it is commonly thought that at least in the finite case  explanations of this kind can be given successfully. But if   there is no rationally compelling way of assigning PSL credences, then even in the finite case we are not obliged to assign credences over universes in any particular way.   From this point of view, the only real difference between the infinite and the finite case is that in the finite case there happens to be a particular choice of `measure' (i.e. a way of assigning  credences over worlds) which has a strong intuitive appeal; but the fact that a measure is intuitively appealing  does not make it rationally compelling. Friederich argues that in the infinite case, `\emph{even if some specific measure were established as physically privileged in the context of external inflation, this would not by itself show that this measure should guide our assignment of probabilities}'\cite{friedbook} and I contend that in fact the same goes for the finite case - credences obtained from simply taking ratios of numbers of universes may be the most obvious choice, but that does not mean we are rationally obliged to set our credences that way. 

What does this mean for explanations which rely crucially on these pure self-locating credences? The answer may   depend on the view of  explanation that one adopts. Certainly if one is working with  a  deductive-nomological  or inductive-statistical  approach\cite{sep-scientific-explanation}, explanations which depend on PSL credences look problematic if there is no rationally compelling way of assigning such credences, for that means we will not be able to derive the explanandum either deterministically or statistically  from just a set of  initial conditions plus some laws of nature: we must in addition make use of a special assignation of self-locating credences  which    simply encodes some kind of subjective attitude, such as how much we care about various individuals. It seems doubtful that such a thing is a legitimate ingredient in either a DN or an IS explanation. Similarly, it is hard to see how we could give a satisfactory causal explanation\cite{Dowe2000-DOWPC-2} relying crucially on purely subjective PSL credences. And even in an approach to explanation based on unification\cite{Kitcher1989-KITEUA}, it's not obvious that an explanation can  be considered significantly unifying if it relies on an essentially arbitrary input which does not come from any relevant theory but  which simply stipulates PSL credences  as a subjective attitude. 

On the other hand, it is true  that in certain cases there is a particular assignation of PSL credences which feels intuitively natural, so if our main desideratum for explanations is that they should provide an intuitive  feeling of understanding, perhaps the use of  intuitively natural PSL credences may be acceptable. However, I want to highlight two problems which could follow from taking these `explanations' too seriously.   The first is that if we are satisfied by such explanations,  this may stop us from exploring paths to more physically-grounded explanations, and then we could  miss out on useful insights into physical reality that would follow from such explanations.  The second is that if we are satisfied by such explanations, we may be tempted to employ them in the context of inference to the best explanation - for  example, this often occurs in arguments for the multiverse, where the idea that the existence of  the multiverse would explain the values of certain  fundamental parameters is used to argue that we ought to believe in such a multiverse\cite{friedbook,Azhar2013-AZHPAT}. But if the explanation in question is the `best' explanation only in the sense that it feels intuitive, it's unclear that we are really justified in making strong inferences about existence from it. So although explanations using PSL credences may in certain circumstances be acceptable, if our primary focus is on achieving an intuitive feeling of understanding, we should be careful about using such explanations to motivate any stronger scientific conclusions. 

With that said, it should be emphasized that the concerns I have raised in this article about the status of PSL credences do not necessarily impugn all kinds of anthropic reasoning. For example, Carter's original formulation of the anthropic principle states that `\emph{what we can expect to observe must be restricted by the conditions necessary for our presence as observers}'\cite{longair2013confrontation} and this principle does not appear to depend on the existence of  rationally compelling PSL credence distributions. Rather it simply mandates that as a matter of certainty we must  find ourselves in a universe belonging to the set of universes obeying various conditions  - and as discussed in section \ref{certainty}, certainty in this sense need not be understood as merely a limiting case of PSL credence, since it can be analysed as a third-person claim about the compatibility of certain experiences with certain physical circumstances. So although the concerns I have raised about PSL credences suggest that there may be no further fact of the matter about   how we ought to assign credences over universes within the relevant set,  that does not undermine the objectivity of the original statement that we will definitely find ourselves somewhere in this set, and thus certain kinds of anthropic reasoning will remain available even if there are no rationally compelling assignments of PSL  credences. 
 
\section{Conclusion}

A famous dilemma for self-locating credences involves a  `presumptuous philosopher' \cite{Bostrom2002-BOSABO-2} who uses self-locating credences to conclude that a certain theory of physics must be right, and then  advises the physicists that they need not even bother performing the experiment to distinguish between two competing theories. Bostrom's response to this scenario is to criticize the particular way in which this philosopher arrived at these self-locating credences\cite{Bostrom2002-BOSABO-2}. But the arguments given this article suggest a much more general response:  it is `presumptuous' under   any circumstances to use PSL credences to arrive at substantive conclusions about physics or the content of reality, because there is no rationally compelling way to assign PSL credences, so such credences  are not a suitable basis for scientific reasoning. Thus in fact, if we take it that the credences involved in the presumptuous philosopher case should be understood as PSL credences, then no matter how the philosopher arrives at them he is in the wrong for trying to answer this question by appeal to PSL  credences alone!

More generally, if the thesis of this article is true, then we should not  expect to resolve substantive scientific questions using PSL credences, and this has important consequences for reasoning around multiverses of various kinds, the simulation hypothesis, Boltzmann brains and so on. Of course, it is entirely possible that much of this reasoning can be rewritten in such a way as to explicitly invoke  SSL credences rather than PSL credences - in this article I have not attempted to determine whether or not this can be done. But even if such rewriting is possible, simply showing explicitly how to achieve it would surely in and of itself represent a major step forward in our understanding of the epistemology of such scenarios. Thus distinguishing clearly between PSL and SSL credences in these applications may help demarcate scientific and unscientific applications of the notion of self-location, which has the potential to significantly clarify ongoing discussions on these topics.

\section{Acknowledgements}

Thanks to Kelvin McQueen for discussions which inspired the writing of this paper, and to Chris Meacham for very helpful comments on a draft of the paper. Thanks also to anonymous referees for their advice!

\end{document}